\documentclass[twocolumn,superscriptaddress,showpacs,floatfix,aps,prl]{revtex4-1}
\usepackage{graphicx}
\usepackage{dcolumn}
\usepackage{amsmath}
\usepackage{amssymb}
\usepackage{amsbsy}
\usepackage{textcomp}
\usepackage{units}
\usepackage{color}
\usepackage{units}
\usepackage{ulem}
\usepackage{pdfpages}
\newcommand{\twotwofour}{2$\times$2$\times$4}
\newcommand{\threethreefour}{3$\times$3$\times$4}

\begin{document}

\title{Magnetism, Spin Texture and In-Gap States: Atomic Specialization at the Surface  of Oxygen-Deficient SrTiO$_3$}

\author{Michaela Altmeyer}
\affiliation{Institut f\"ur Theoretische Physik, Goethe-Universt{\"a}t Frankfurt, 60438 Frankfurt am Main, Germany}
\email{altmeyer@itp.uni-frankfurt.de}

\author{Harald O. Jeschke}
\affiliation{Institut f\"ur Theoretische Physik, Goethe-Universt{\"a}t Frankfurt,  60438 Frankfurt am Main, Germany}

\author{Oliver Hijano-Cubelos}
\affiliation{Laboratoire de Physique des Solides, Bat 510, Universit\'e Paris-Sud,  91405 Orsay, France}

\author{Cyril Martins}
\affiliation{Laboratoire de Physique des Solides, Bat 510, Universit\'e Paris-Sud,  91405 Orsay, France}

\author{Frank Lechermann}
\affiliation{I. Institut f\"ur Theoretische Physik,
Universit\"at Hamburg, 20355 Hamburg, Germany}
\affiliation{Institut f. Keramische Hochleistungswerkstoffe, TU Hamburg-Harburg, D-21073 Hamburg, Germany}

\author{Klaus Koepernik}
\affiliation{IFW Dresden, Helmholtzstra{\ss}e 20, 01069 Dresden, Germany}

\author{Andr\'es F. Santander-Syro}
\affiliation{CSNSM, Univ. Paris-Sud, CNRS/IN2P3, Universit\'e Paris-Saclay, B\^at. 104 et 108, 91405 Orsay, France}

\author{Marcelo J. Rozenberg}
\affiliation{Laboratoire de Physique des Solides, Bat 510, Universit\'e Paris-Sud, 91405 Orsay, France}

\author{Roser Valent\'\i}
\affiliation{Institut f\"ur Theoretische Physik, Goethe-Universt{\"a}t Frankfurt, 60438 Frankfurt am Main, Germany}

\author{Marc Gabay}
\affiliation{Laboratoire de Physique des Solides, Bat 510, Universit\'e Paris-Sud, 91405 Orsay, France}

\begin{abstract}

  Motivated by recent spin- and angular-resolved photoemission (SARPES)
  measurements of the two-dimensional electronic states
  confined near the (001) surface of oxygen-deficient SrTiO$_3$, we
  explore their spin structure by means of {\it ab initio} density
  functional theory (DFT) calculations of slabs.  Relativistic
  nonmagnetic DFT calculations display Rashba-like spin winding with
  a splitting of a few meV and when surface magnetism on the Ti ions
  is included, bands become spin-split with an energy difference $\sim100$ meV at the $\Gamma$ point, consistent with SARPES findings.
  While magnetism tends to suppress the effects of the relativistic
  Rashba interaction, signatures of it are still clearly visible in
  terms of complex spin textures. Furthermore, we observe an {\it
    atomic specialization} phenomenon, namely, two types of electronic
  contributions: one is from Ti atoms neighboring the oxygen vacancies
  that acquire rather large magnetic moments and mostly create in-gap
  states; another comes from the partly polarized $t_{2g}$ itinerant
  electrons of Ti atoms lying further away from the oxygen vacancy,
  which form the two-dimensional electron system and are responsible
  for the Rashba spin winding and the spin splitting at the Fermi
  surface.

\end{abstract}

\maketitle

{\it Introduction.-} Transition metal oxides constitute a major topic
of interest in the scientific community as these materials are endowed
with a broad range of significant functionalities, ranging from
ferroelectricity to metal-insulator transitions as well as from
magnetism to superconductivity. Many of these compounds exhibit
structural instabilities, strong electronic correlations, and complex
phase diagrams with competing ground states.  Artificial structures of
transition metal oxides therefore seem ideal to explore interfacial
effects that could possibly lead to new phases.  In this respect, the
observation a decade ago~\cite{Ohtomo2004,Thiel2006} of a
two-dimensional electronic system (2DES) at the interface between the
wide band-gap insulators LaAlO$_3$ (LAO) and SrTiO$_3$ (STO) has
attracted a considerable amount of attention. It was found that the
2DES hosts gate-tunable insulator to metal, insulator to
superconductor transitions, magnetism~\cite{Bert2011}, and a large
interfacial spin-orbit effect~\cite{Sulpizio2014}.  The mechanisms
responsible for these special properties are still under debate.
Questions include the intrinsic (i.e., electronic reconstruction)
versus extrinsic (e.g., oxygen vacancies)
mechanism~\cite{Vonk2012,Cancellieri2011,Berner2013} responsible for
the formation of the 2DES, and also the role and spatial distribution
of the various $d$ orbitals that could contribute selectively to a
specific charge or spin property.

Angular-resolved photoemission (ARPES) measurements revealed the
existence of a 2DES with similar features to those seen at the LAO-STO
interface at the {\it bare surfaces} of several insulating perovskite
oxide crystals, and among them (001) oriented
STO~\cite{Santander-Syro2011,Baumberger2011}. In this case, the
carriers originate from oxygen vacancies.  These vacancies are likely
created during the sample preparation process and also when the sample
is illuminated during the measurement~\cite{Walker2014}.  Spin
resolved ARPES (SARPES) measurements of the 2DES at the (001) oriented STO
surface~\cite{Santander-Syro2014} have highlighted the existence of
sizable Rashba-like spin textures along with a large energy splitting
that has been interpreted as a signature of ferromagnetism. Their
simultaneous occurrence is puzzling since the two effects \textit{a
  priori} compete with each other and estimates of the respective
energy scales give about 100~meV for magnetism and a few meV for
surface spin-orbit coupling.

A number of density functional theory (DFT) studies on
oxygen-deficient bulk STO~\cite{Cuong2007,Hou2010,Lopez2015} found
magnetism and in-gap bound states below the conduction band.  The
reported location of the bound states and the size of the spin
polarization depend strongly on the computational details.  Recent
nonmagnetic DFT calculations for slab
geometries~\cite{Shen2012,Jeschke2015} found the formation
of a 2DES at the surface, in addition
to  
in-gap states. Effects of spin orbit coupling~\cite{Zhong2013,Khalsa2013} and
correlation~\cite{Lin2013} have also been studied at the model level.

In the present work, we investigate via first principles DFT the spin
textures and magnetism of the 2DES states at the (001) oriented
surface of oxygen-deficient STO.
 As the preexisting literature on this problem
is already large, let us emphasize that the new results reported here
are relevant for the interpretation of the following experimental
facts:  (i) spin ARPES
appears to highlight the occurrence of both Rashba-like textures and
ferromagnetism~\cite{Santander-Syro2014},
and (ii) ARPES spectra pertaining to the 2DES at interfaces and also at
surfaces all seem to appear concomitant with a universal
nondispersive feature at an energy of about 1.3~eV below the Fermi
energy~\cite{Aiura2002,Courths1980,Kim2009}.
Our DFT calculations are accompanied by tight-binding model considerations.

\begin{figure}[htb]
\centering
\includegraphics[width=\columnwidth]{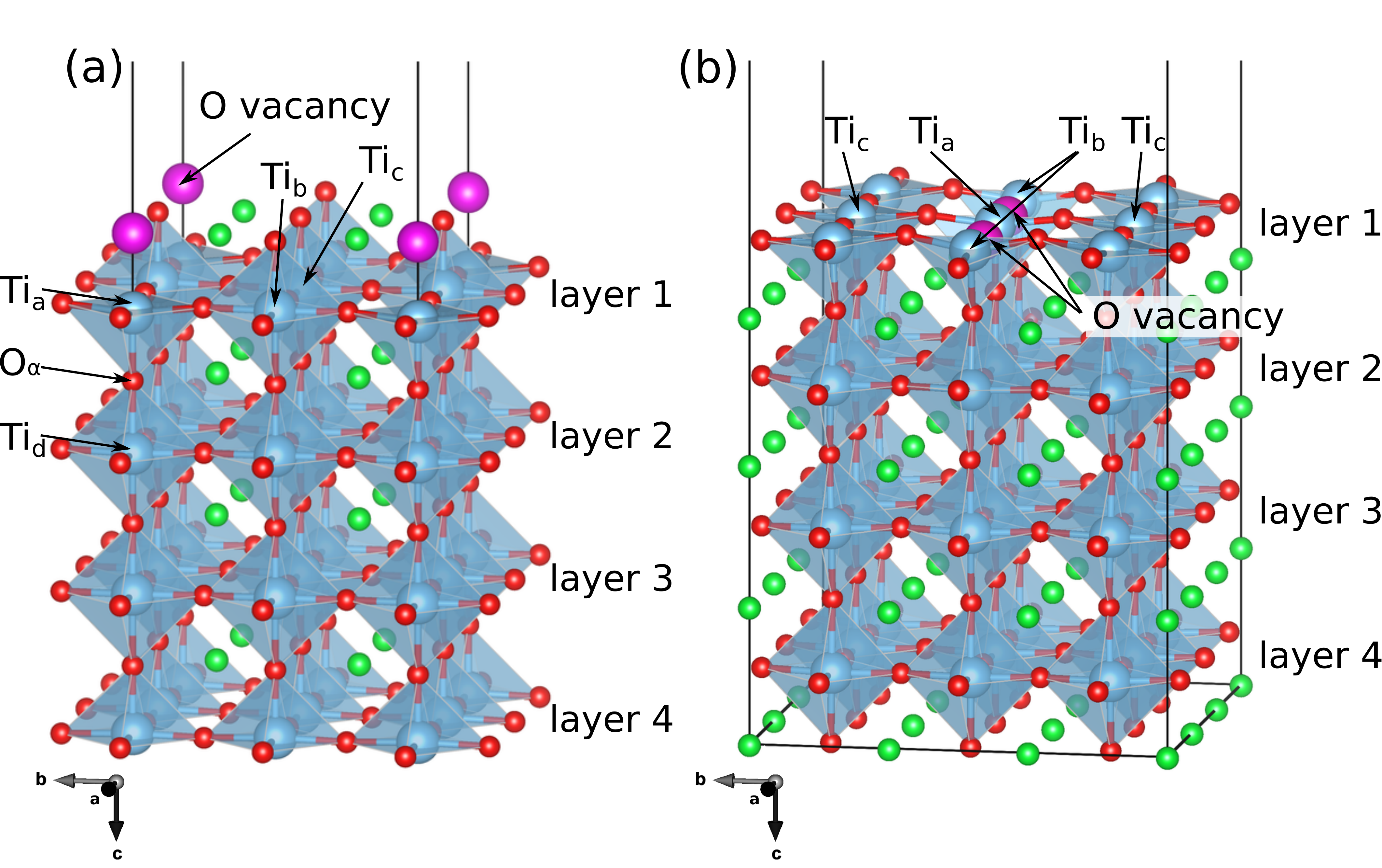}
\caption{(a) SrO-terminated {\twotwofour} slab with one oxygen vacancy
  at the topmost level. O$_\alpha$ denotes the position where a second
  vacancy is introduced for divacancy {\twotwofour} calculations (see
  text).  (b) TiO$_2$-terminated {\threethreefour} slab with two
  vertically positioned oxygen vacancies.  Other slab geometries are
  included in the Supplemental Material.}
\label{fig:structures}
\end{figure}

{\it Methods.-} 
In order to extract the basic trends in the electronic properties we considered
three representative slabs of (001) oriented STO with different oxygen
vacancy concentrations and slab terminations in our DFT calculations:
a single vacancy at the topmost level of a
SrO-terminated {\twotwofour} slab [Fig.~\ref{fig:structures}~(a)], a
vertical divacancy at the topmost level of a SrO-terminated
{\twotwofour} slab [Fig.~\ref{fig:structures}~(a)], and a horizontal
divacancy located at the topmost level of a TiO$_2$-terminated
{\threethreefour} slab [Fig.~\ref{fig:structures}~(b), \cite{comment1}].  These slabs
are chosen as test cases for investigating the effects of surface
reconstruction, in-gap states, Rashba spin-orbit coupling and
magnetism. We have made sure with a larger set of slabs (not shown)
that all observations discussed below do not crucially depend on TiO$_2$
versus SrO termination. In all slabs we included a vacuum layer of at least
20~{\AA} to avoid any spurious interactions between the periodic
images.
To our knowledge, no measurements of vacancy densities have been performed 
so far; therefore we have chosen our test slabs, such that the surface vacancy induced
carrier densities are comparable to the experimental observations of 
$n_{2D}\approx 0.25-0.33 $ e$^-$/$a^2$ (see f.i. 
\cite{ commentbulkvacancy, Santander-Syro2011, Baumberger2011}). 
Every vacancy nominally releases two electrons. Because of the in-gap state
not all the additional electrons are contributing to the 2DES, so that we can estimate a 
required density of about one vacancy per $4$ unit cells. 
Our mono- and divacancy  {\twotwofour} and {\threethreefour} slabs are therefore
compatible with the experimental estimates.

\begin{figure*}[htb]
\centering
\includegraphics[width=0.95\textwidth]{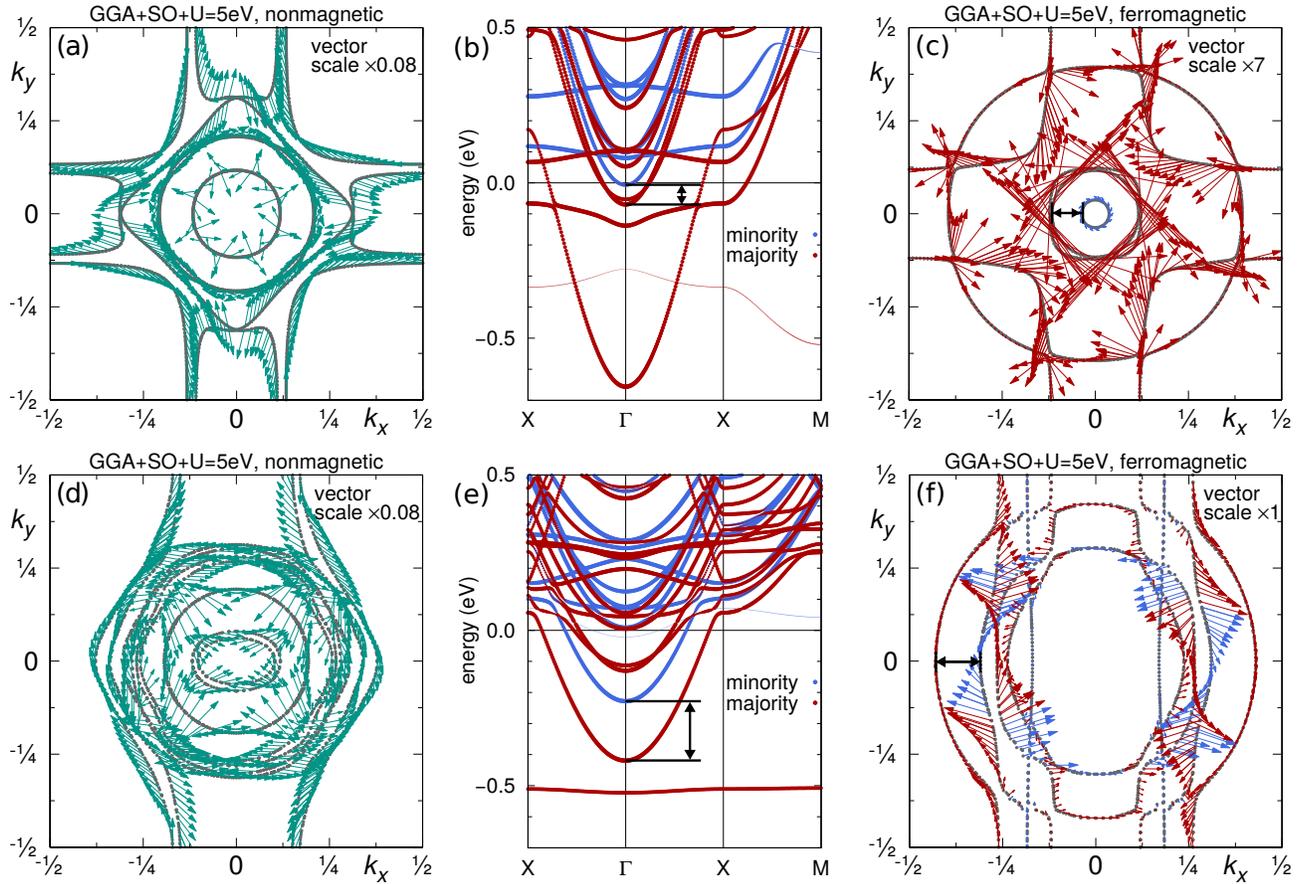}
\caption{Spin-textures and spin-polarized band structures for the
  monovacancy {\twotwofour} slab (a)-(c) and the
  divacancy {\threethreefour} slab (d)-(f). (a) and (d) are nonmagnetic
  GGA+SO+U calculations with $U=5$~eV, and (b), (c), (e), and (f) are
  ferromagnetic GGA+SO+U calculations with $m \parallel
  \hat{z}$. Note that the large magnetic moments along $\hat{z}$ can
  be inferred from the exchange splitting in (b) and (e); the shown
  spin textures are projections on the $xy$ plane. Reciprocal space
  units are $2\pi/(2a)=0.805$~{\AA}$^{-1}$ for (a) and (c), and $2\pi/(3a)=0.536$~{\AA}$^{-1}$ for
  (d) and (f), where $a$ is the bulk STO lattice parameter. The different vector scales are chosen to enhance the 
  visibility of the spin windings. }
\label{fig:spintexture}
\end{figure*}

In order to account for possible surface reconstructions, the internal
coordinates of the slabs were relaxed with the projector-augmented
wave basis~\cite{Bloechl1994a} as implemented in
VASP~\cite{Kresse1996,Hafner2008}. We used the generalized-gradient approximation
(GGA)~\cite{Perdew1996} in the Dudarev~\cite{Dudarev1998} GGA+U scheme
as described in Ref.~\onlinecite{Shen2012}.  The electronic structure
was analyzed with the all-electron full-potential local orbital ({\sc
  FPLO})~\cite{FPLO} method and GGA+U
functional~\cite{Liechtenstein1995}. Checks were also performed with
the  linearized augmented plane wave method
as implemented in Wien2k~\cite{Blaha2001}.  Spin textures for the
various slabs were obtained from full relativistic calculations using the full-potential local orbital GGA+SO+U method with a newly implemented subroutine.

{\it Results and discussion.-} We start with the analysis of
spin textures in the absence of magnetism.  In
Figs.~\ref{fig:spintexture}(a) and \ref{fig:spintexture}(d) we show
the spin textures (projections of the spin polarization vectors onto 
the $xy$ plane) at the Fermi surface ($k_z=0$) obtained for
nonmagnetic ground states in GGA+SO+U calculations for the monovacancy 
{\twotwofour} and the divacancy {\threethreefour} slabs,
respectively (results for the divacancy {\twotwofour} are shown in the
Supplemental Material).  We used typical values for the parameters ($U=5$~eV and
$J_{\rm H}=0.64$~eV on Ti $3d$ orbitals~\cite{Okamoto2006}). In all cases,
every two bands show a small energy splitting of a few meV 
 due to the spin-orbit interaction. Spins at the Fermi
surface ($k_z = 0$) are fully polarized in the $xy$ plane, oppositely
oriented in the split bands. On some of the bands the spins are
pointing clock- and anticlockwise around $\Gamma$, which is a clear
signature of the relativistic Rashba effect. On other bands we
additionally notice a texture of rotating spins, born out
of the more complex interplay between spin and orbital degrees of
freedom~\cite{Walker2014,King2014}.  The rather small size of the
spin splitting contrasts with the large value reported in recent
SARPES experiments~\cite{Santander-Syro2014}.

\begin{figure}[htb]
\centering
\includegraphics[width=0.98\columnwidth]{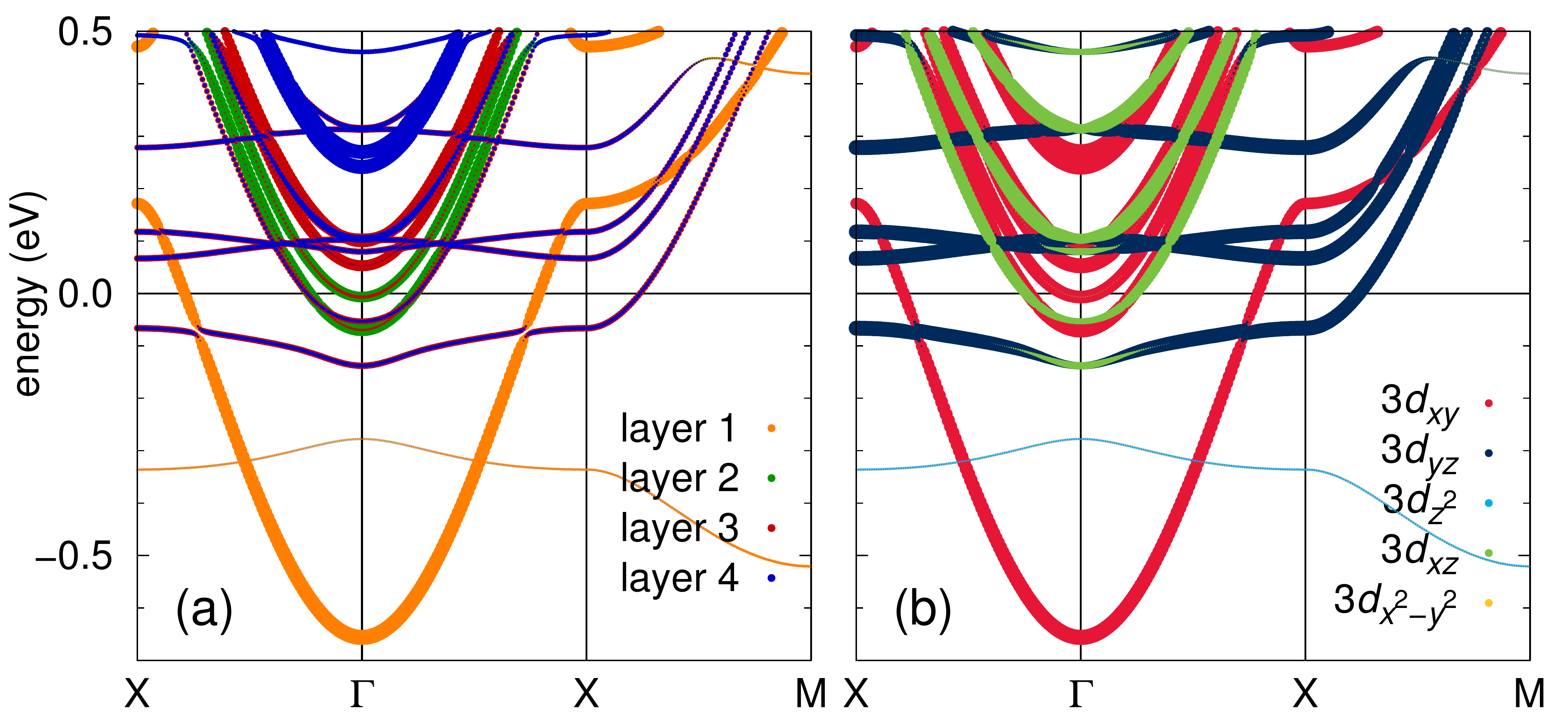}
\caption{Ferromagnetic GGA+SO+U Ti $3d$ band structure for the monovacancy
  {\twotwofour} slab with $m \parallel \hat{z}$. (a) Layer resolved. 
  (b) Orbitally resolved. Note that the
  thickness of lines is proportional to the strength of the $3d$
  character on the bands (the heavy band at $-0.4$~eV is strongly 
  hybridized with the Ti $4s$ and $4p$ orbitals, see also the main text).}
\label{fig:224bandstructure}
\end{figure}

\begin{figure}[htb]
\centering
\includegraphics[width=0.9\columnwidth]{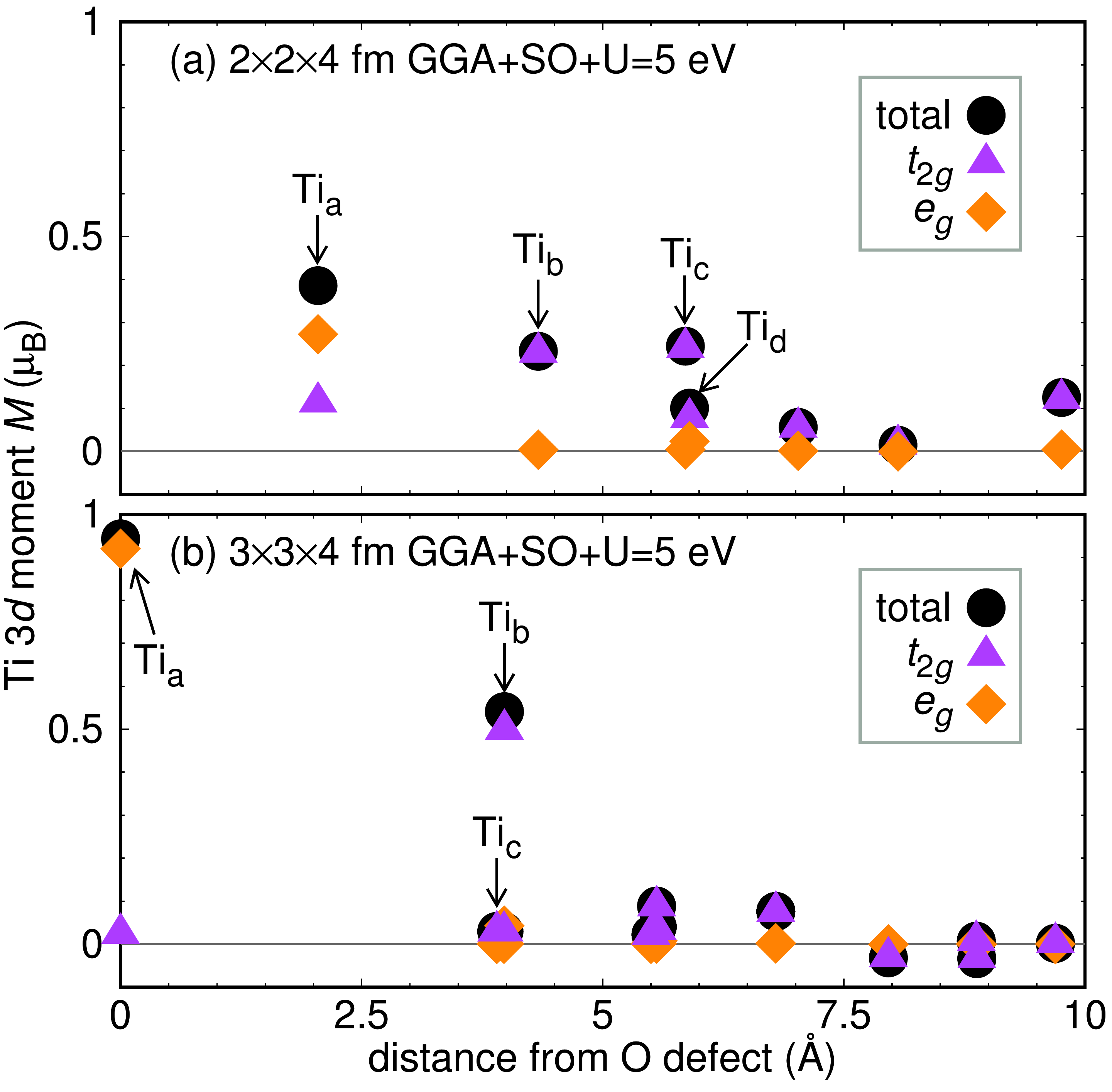}
\caption{$t_{2g}$-$e_g$ resolved magnetic moments in GGA+SO+U=5~eV for
  the  (a) monovacancy {\twotwofour} and (b) divacancy {\threethreefour} slab in the
  ferromagnetic $m \parallel \hat{z}$ setup. Note that the moments shown do not contain 
  hybridization contributions.}
\label{fig:224mm}
\end{figure}

Next, we consider solutions with ferromagnetic order and spin-orbit
interactions. The ferromagnetic solution is indeed the ground state of
the systems we consider~\cite{comment2}.
Figure~\ref{fig:spintexture}~(b) displays the spin-projected
band structure obtained from spin-polarized GGA+SO+U calculations for
the monovacancy {\twotwofour} slab (similarly, results for the
divacancy {\twotwofour} slab are shown in the Supplemental Material).  We adopt
the magnetic moment quantization axis along $z$ but below we also
discuss the case of a quantization along $x$.  The size of the
magnetic splitting can be inferred from the black arrows connecting
the majority and minority spin bands. For a comparison to the
experiment we have to consider the splitting of the light bands of
$d_{xy}$ character, as heavy bands have been silenced in the
measurements~\cite{Santander-Syro2014}.  The energy separation at the
$\Gamma$ point of the two spin-split $d_{xy}$ bands originating from
Ti$_{\rm d}$ [Fig.~\ref{fig:structures}(a)] is of the order of
$60$~meV and in qualitative agreement with the experimental
data~\cite{Santander-Syro2014} ($\Delta E \approx
100$~meV).

In order to identify the microscopic origin of the peculiar electronic
and magnetic features described above we plot in
Fig.~\ref{fig:224bandstructure} the layer- and orbital-resolved
GGA+SO+U band structure near $E_F$ for the monovacancy {\twotwofour}
slab  and in Fig.~\ref{fig:224mm}~(a) we display the Ti
$t_{2g}-e_g$ resolved magnetic moments as a function of the distance
between Ti and O vacancy.  We find that (i) the magnetic splitting of
the light $d_{xy}$ bands at the Fermi level is caused by itinerant
electrons belonging to Ti located not in the immediate vicinity of the
oxygen vacancy [e.g. Ti$_{\rm d}$ in Fig.~\ref{fig:structures}~(a)]
with Ti $3d$ magnetic moments of the order of 0.1 $\mu_B$ [see
Fig.~\ref{fig:224mm}~(a)].  (ii) Ti atoms neighboring the oxygen
vacancy in the uppermost layer [Ti$_{\rm a}$, Ti$_{\rm b}$, Ti$_{\rm
  c}$ in Fig.~\ref{fig:structures}~(a)] have the largest magnetic
moment [Fig.~\ref{fig:224mm}~(a)] and are mostly responsible for the
heavy bands (Ti$_{\rm a}$) and occupied states at higher binding energies [Ti$_{\rm b}$, Ti$_{\rm c}$, see Fig.~\ref{fig:224bandstructure}]. In particular, we observe an in-gap band of Ti $e_g$ ($d_{z^2}$) character hybridizing with Ti $4s$ and $4p$ corresponding to Ti$_{\rm a}$. It sits at -0.4~eV in the single
vacancy case and is shifted to about -1~eV in the 2$\times$2$\times$4 divacancy
case (see the Supplemental Material).  The position of this band depends on the parameters $U$
and $J_{\rm H}$ chosen for the GGA+U calculations and on the concentration
and position of vacancies in the slab (see Supplemental Material and
Ref.~\onlinecite{Jeschke2015}).

Interestingly, already with this minimal slab, we find a phenomenon of
{\it atomic specialization}; i.e.,  there are two types of electronic
contributions to magnetism: one from Ti atoms neighboring the oxygen
vacancy that acquire rather large magnetic moments and are mostly
located below the Fermi surface inducing in-gap states, and another,
from those Ti atoms lying further away from the oxygen vacancies, that
correspond to polarized $t_{2g}$ itinerant electrons with small
magnetic moments, which are responsible for the Rashba spin winding
and the spin splitting at the Fermi surface. These remarkable effects
will be even more pronounced in larger slabs.

We observe here the same phenomenon of atomic specialization as in the
smaller {\twotwofour} slab; Ti atoms neighboring the oxygen vacancies
[Ti$_{\rm a}$ and Ti$_{\rm b}$ in Fig.~\ref{fig:structures}~(b)]
acquire large magnetic moments [see Fig.~\ref{fig:224mm}~(b)] and are
responsible for the in-gap states located at higher binding energies,
while Ti atoms lying further away from the oxygen vacancy (Ti$_{\rm
  c}$ and beyond) contribute to the 2DES with itinerant electrons
carrying small magnetic moments, which are responsible for the Rashba
spin winding and the spin splitting at $E_F$~\cite{comment3}.

Further insight into the atomic dichotomy can be gained through
tight-binding cluster diagonalization (see Supplemental Material) of structures
with various configurations of vacancies. In order to monitor the
formation of the 2DES conduction band states and of the in-gap states,
we adiabatically turn on the energy contributions that represent the
effect of introducing a vacancy into the cluster.  Features seen in
DFT are qualitatively reproduced.

Based on our above discussion of each individual
effect, we may infer that spin textures and spin splitting compete
with each other in the $t_{2g}$ bands.  Fig.~\ref{fig:spintexture}~(c)
displays the spin texture at $k_z=0$ obtained from spin-polarized
GGA+SO+U calculations for the monovacancy {\twotwofour} slab.
The spin texture shows signs of the Rashba winding but it is less
pronounced than in the nonmagnetic case.  Taking a closer look at the
inner pockets in Fig.~\ref{fig:spintexture}~(c) (blue and red circles
centered at $\Gamma$) corresponding to the spin up and spin down
projections [compare Fig.~\ref{fig:spintexture}~(b)], we observe a
significant Fermi momentum shift of the bands ($\sim 0.1
\textnormal{\AA}^{-1}$), which is of the same order of magnitude as
the one observed in SARPES experiments~\cite{Santander-Syro2014}.  As
can be inferred from the now very small in-plane spin component
($P_\parallel \leq 0.04$), ferromagnetism is dominating the
arrangement of the spins but the spin winding is still visible. The
same features are observed for the divacancy 3$\times$3$\times$4 slab
[Fig.~\ref{fig:spintexture}~(f)], confirming the general validity of
the results. 
We have also checked the robustness of the results with respect to the choice of $U$
in the GGA+U functional and find that the variation of $U$ introduces
only quantitative changes in the 
ferromagnetic calculations (see Supplemental Material).

Scanning superconducting quantum interference device
measurements~\cite{Bert2011} on LAO/STO interfaces observed a
preference of in-plane magnetic moments, which are ascribed to shape
anisotropy.  As this is also expected for pure STO surfaces, we
additionally performed calculations with the magnetization axis along
$x$; in this case the spins are aligned in plane and the Rashba
interaction is unable to rotate the spins to achieve a sign change for
opposite $k$ points. However, a canting of the spin polarization
vectors away from the magnetic axis originating from the Rashba
coupling can be still identified, in agreement with previous
theoretical considerations~\cite{Barnes2014}.  Our results would be
compatible with SARPES spectra, which see both an almost pure Rashba
spin texture and a full in-plane polarization, if we assume that the
measurements detect a signal from a large number of ferromagnetic
domains at once.  In this case the effect of magnetism on the
alignment of the spin polarization vectors is averaged out and the
measured spin texture is completely determined by spin-orbit coupling
effects, restoring the sign change for opposite $k$ points and
implying that the main features are independent of the magnetic
quantization axis.

{\it Conclusions.-} By performing full relativistic nonmagnetic and
magnetic density functional theory calculations in the framework of the
GGA+SO+U functional on representative oxygen deficient SrTiO$_3$ slabs, we find
the magnetic state to be the ground state and we observe clear
signatures of {\it atomic specialization} of the electronic and
magnetic contributions.  Ti atoms neighboring the oxygen vacancies
create $e_g$ localized wave functions with large magnetic moments and
are responsible for the presence of in-gap states at energies around
-0.5 to -1~eV. The position of the in-gap states is influenced by the
slab termination, the depth of the oxygen vacancy below the surface,
and by possible oxygen clustering.  On the other hand, Ti atoms lying
further away from the oxygen vacancy contribute with polarized
$t_{2g}$ itinerant electrons to the conducting 2DES and are
responsible for the Rashba spin winding and the spin splitting at the
Fermi surface observed in SARPES.  Our calculations show that
magnetism masks the Rashba effect by increasing the spin splitting of
the $t_{2g}$ orbitals and by modifying the individual spin orientation
but it does not eliminate spin winding.  Considering that an averaging
of inhomogeneities near the surface of the measured sample is to be
expected (e.g., a sea of the 2DES interspersed with islands of magnetism
perhaps mirroring a mixture of TiO$_2$ and SrO in the termination
layer), SARPES measurements are explained by our calculations as the
combined effect of the Rashba effect and magnetism.

\begin{acknowledgments}

We would like to thank Vladislav Borisov for performing test
calculations with Quantum Espresso. We especially thank Domenico di
Sante, Silvia Picozzi, Michael Sing and Ralph Claessen for useful
discussions. M.A., H.O.J. and R.V. gratefully acknowledge the Deutsche
Forschungsgemeinschaft (DFG) for financial support through Grants
SFB/TR 49, FOR 1346. M.A. and R.V. were partially supported by the
Kavli Institute for Theoretical Physics at the University of
California, Santa Barbara under National Science Foundation Grant No.
PHY11-25915. A.F.S.-S. and M.G. acknowledge support from the Institut Universitaire de France and from the French National Research Agency (ANR) (Project LACUNES No. ANR-13-BS04-0006-01).
 
\end{acknowledgments}


\begin{thebibliography}{99}
\bibitem{comment1} In the {\threethreefour} slab [Fig.~\ref{fig:structures}(b)] the loss
of symmetry due to the presence of the vacancies induces a ladder of
bands at $E_F$ corresponding to inequivalent Ti atoms
[Fig.~\ref{fig:spintexture}~(f)] and a breaking of the $xy$ symmetry
as also observed in the Fermi surface topography
[Fig.~\ref{fig:spintexture}~(d)].  In experiment, such an asymmetry is
not observed and is here a consequence of considering an individual
divacancy realization. This asymmetry is expected to decrease when
larger slab geometries and a disorder average are considered, which is
beyond our computational possibilities.
\bibitem{comment2} At the GGA+SO+U=5~eV level, the ferromagnetic
  solution lowers the energy by 14.9~meV/f.u. for a single O vacancy
  in the {\twotwofour} slab, and by 7.3~meV/f.u. for a divacancy in
  the {\threethreefour} slab; the difference in stabilization energy
  arises from the respective total moments of 0.125~$\mu_{\rm
    B}$/f.u. and 0.077~$\mu_{\rm B}$/f.u. which are averaged over the
  16 and 36 formula units of the two slabs.
\end{thebibliography}
\end{document}